\documentclass{sig-alternate-10pt}
\pdfoutput=1

\makeatletter
\newcommand\gobblepars{%
    \@ifnextchar\par%
        {\expandafter\gobblepars\@gobble}%
        {}}
\makeatother

\renewcommand{\paragraph}[1]{\smallskip\noindent\textbf{{#1}.}\ \ \gobblepars}

\usepackage{amsmath} %
\usepackage{ellipsis} %
\usepackage{url} %
\usepackage{epsfig,color} %
\usepackage{stmaryrd} %
\usepackage{array} %
\usepackage{balance}

\newcommand{\degen}[1]%
{\delta({#1})}

\newcommand*{\given}{\mathop{{|}}}
\newcommand{\msf}{\mathsf}

\DeclareMathOperator*{\argmin}{argmin}

\newcommand{\op}{\msf{op}}

\newcommand{\meas}{\msf{meas}}
\newcommand{\store}{\msf{store}}

\newcommand{\cost}{\msf{cost}}
\newcommand{\imp}{\msf{imp}}

\title{On Modeling the Costs of Censorship\thanks{We gratefully acknowledge funding support from the Freedom 2 Connect Foundation, Intel, the National Science Foundation (grants 1237265 and 1223717), the Open Technology Fund, the TRUST Science and Technology Center (NSF grant 0424422), the US Department of State Bureau of Democracy, Human Rights, and Labor.  The opinions in this paper are those of the authors and do not necessarily reflect the opinions of any funding sponsor or the United States Government.}}
\date{}
\author{Michael Carl Tschantz\\ UC Berkeley 
\and Sadia Afroz\\ UC Berkeley
\and Vern Paxson\\ International Computer Science Institute\\ and UC Berkeley
\and J. D. Tygar\\ UC Berkeley}

\begin{document}
\maketitle

\begin{abstract}
We argue that the evaluation of censorship evasion tools should
depend upon economic models of censorship.  We illustrate our position
with a simple model of the costs of censorship.  We show how this model
makes suggestions for how to evade censorship.  In particular, from it,
we develop evaluation criteria.  We examine how our criteria compare to
the traditional methods of evaluation employed in prior works.
\end{abstract}

\section{Introduction}

\paragraph{Motivation}
In response to government censorship of the Internet, activists
and researchers have deployed numerous censorship evasion
tools~\cite{callanan11freedomhouse}.  Censors, in turn,
have developed approaches to counter these evasion tools
by, for example, blocking all Internet traffic produced from a given
tool~\cite{roberts11berkman,kelly13freedomhouse}.  Researchers have responded
by proposing
numerous improvements intended to neutralize such blocking (e.g.,
\cite{wiley11dust,mohajeri2012skypemorph,weinberg2012stegotorus,winter2013scramblesuit}).
Due to limited resources, the developers of evasion tools cannot implement
and deploy all of these; they need criteria for selecting the most promising.

While
the evaluation sections of research papers provide some insight into the
promise of each proposal, each paper employs its own evaluation
methodology, typically selected with the capabilities of the tool in mind,
making cross-tool comparisons difficult.  Furthermore, determining how
well the evaluation predicts real-world performance often poses difficulties.
Prototyped tools meeting evaluation criteria are sometimes easily broken
using methods unconsidered by the evaluation~\cite{houmansadr13sp}.

In this paper, we provide a preliminary examination of a different framework
for evaluation with hopes of generating interest in evaluation issues.
We propose augmenting prior tool-specific
methods of evaluation with a new methodology for creating evaluation
criteria and interpreting results.  We start from the premise that censors
and evaders engage in an ongoing arms race whose ebb and flow is largely
determined by economic concerns.  Thus,
we argue we should evaluate tools on their
promise to drive up the costs of the censor while remaining inexpensive to
implement.

From this prospective, tool-specific evaluations become important
evidence that a tool could drive up a censor's costs.  Our
perspective addresses
the abovementioned shortcoming of prior evaluations
in two ways.  First, by examining total cost, we remind the evaluator that
every aspect of the traffic produced by the evasion tool matters, not
simply those considered by its designer.  We hope this universal view will
encourage designers to widen their focus and catch the often simple
attacks that foiled past approaches.

Second, by potentially providing a numerical score, cost can reflect
a more quantitative measure
than seeing whether a tool can be broken by any means, a standard more
appropriate when a clear winner is possible (e.g., cryptographic protocols)
than an arms race.

\paragraph{Overview}
After motivating our arms-race view of censorship and the need for considering
costs,
we turn our attention to illustrating the use of our methodology.  Our
illustration is preliminary and focuses on only one side of the equation,
the costs to the censor, leaving the evader's costs to future work.

Our illustration starts with a simple model of the censor's costs.  The
model emphasizes the ability of the censor to employ any feature of network
traffic, not just those on which the tool is evaluated.
Given the paucity of information
regarding the budgets of real censors, our model must leave the actual costs as
unknown parameters.  Thus, we cannot use our model to predict actual costs.
Despite this limitation, we use it to reason about whether a particular
design choice increases costs (by some undetermined amount).

We find that these qualitative results suggest economically motivated
evaluation criteria for evasion tools.  Informally, we judge a tool by
the number of \emph{inexpensive} features it obfuscates.  We estimate the expense of a feature using surrogates for the actual costs faced by the censor.

We examine three prior approaches under these criteria and find that
each is narrowly focused.  We also use our model to explain the
effectiveness of \emph{active manipulation}, attacks during which the censor
interacts with an evasion tool, rather than just passively watching its traffic.  We conclude that, in the case of a blacklisting censor, \emph{looking different from known disallowed traffic is a better choice than mimicking allowed protocols.}
We end with a discussion of open questions.

\section{Prior Work}

Some previous research considered cost of a censor as a result of different circumvention methods. Houmansadr et al.\ determined that evading decoy routing would increase a censor's cost in terms of network latency and path length~\cite{houmansadr2014no}. Elahi et al.\ proposed the CORDON taxonomy that divides different censorship evasion strategies into six types according to their effects on a censor~\cite{elahi2012cordon}. We model costs of the censor for detecting traffic obfuscation tools.

Roberts et al.\ evaluated tools by testing whether they work in various countries~\cite{roberts11berkman}.
 Callanan et al.\ used a combination of in-laboratory tests and user surveys to determine the usability, performance, and security characteristics of a variety of tools~\cite{callanan11freedomhouse}.  By using common methods on tools in their environment, these researchers were able to compare the current success of deployed tools.  
However, 
researchers and developers need criteria applicable to undeployed tools.  They must carefully select which proposals to develop and deploy due the costs associated with such efforts.  Also, criteria based on surveys include factors other than the technical merits of an approach.  For example, a tool may have high adoption due to having first-mover advantage or popular proponents.  Alternatively, a tool might be unblocked despite being easily blocked since the tool is too unpopular to warrant the censor's attention.  We desire evaluation criteria that rate tools upon their technical merits. 

Others have explored absolute characterizations of success.  For example, Pfitzmann and Hansen use \emph{undetectability}, or \emph{unobservability}, to mean that the censor should not be able to determine which Internet users are using the evasion tool and use \emph{unblockability} to mean that the censor should not be able to block the tool's traffic without also blocking a great deal of unrelated traffic~\cite{pfitzmann2010terminology}.  However, perfectly achieving these goals typically leads to unacceptably high performance degradation.  Thus, undetectability and unblockability are only approximated by tools attempting to increase the censor's numbers of false negatives and false positives, respectively.

Dingledine enumerates general properties that make for a good evasion tool~\cite{dingledine10tor}.  We focus on lower-level criteria specific to a tool not being blocked and on the justification of the criteria in terms of an economic model. 

None the particular evaluations used by various tool developers were designed to be applicable to multiple tools.  We discuss them in Section~\ref{sec:eval-tools}, where we compare their evaluations to our own criteria.

Houmansadr et al.\ empirically come to a similar conclusion as we do: tools mimicking allowed protocols are ill-advised~\cite{houmansadr13sp}.  They support their position by showing that censors can easily identify such mimicking tools.  
We explore the issue using a formal model.

\section{The Arms Race}
\label{sec:arms}

When an evader deploys a sufficiently successful evasion tool, an effected censor typically improves its system to catch use of the tool.  Thus, the two sides are engaged in an arms race.  
With this in mind, evasion tools should be designed to slow the censor down and to cost it resources.

To illustrate this arms race and motivate the consideration of costs, we consider the two most popular approaches that evaders take, polymorphism and steganography, and the possible responses of the censors.  Before doing so, we provide background on the censor and evader. We focus on Tor-based evasion tools~\cite{tor} and presume familiarity with Tor's use as such.  

\paragraph{Censors and Evaders}
A censor disallows some subset of messages.  It employs a classifier that examines network traffic and attempts to identify those packets facilitating a disallowed message.  The classifier typically uses hand-crafted signatures that characterize disallowed traffic (a blacklist) or allowed traffic (a whitelist).  We focus on blacklisting censors as they are more common~\cite{kelly13freedomhouse}.  We consider the classifier missing a disallowed message to be a false negative and accidentally blocking an allowed message to be a false positive.

The signatures in the blacklist refer to the value of various \emph{features} of the traffic.  These features can depend upon a single packet, such as IP address or the destination's domain name;  upon distributions over more than one packet, such as the distribution of interpacket arrival times and packet lengths within a packet flow; or upon the sequence of packets within a flow, requiring the keeping of state.  Distributional and stateful features tend to be more costly since they require more storage and computation~\cite{khattak2013towards}.

A censor can identify disallowed traffic either by passive monitoring or by active manipulation.  Whereas monitoring simply watches traffic, active manipulation involves the censor sending traffic to a suspected evader.  For example, the evader can send manipulated requests to the evader to study its reaction.  Active attacks can allow the censor to drive the evader toward more recognizable traffic, which we discuss more in Section~\ref{sec:active}.

We view an evasion tool as a transformation on the network traffic that the censor attempts to classify. 
Since censors typically consider any traffic employing an evasion tool as disallowed, we only consider transformations that alter only disallowed traffic.
(Transformations on allowed traffic could be possible with help from, for example, ISPs, but we leave such considerations to future work.)
Thus, the tool can transform the features examined by the censor in a manner that drives up false negatives, but not false positives.

If an evader drives up the number of false negatives unacceptably high, the censor will respond with a new classifier.  The classifier could be altered to recognize the new values produced by disallowed messages under the old features or to employ new features that remain unobfuscated by the evader.  The classifier may introduce false positives by attempting to block the tool too aggressively.  Thus, tools can indirectly cause false positives.

Let us consider two examples of such transformations with the first motivated by polymorphism and the second by steganography.
We also consider the censor's possible responses to the transformations and how they affect the expenses of the censor.
For simplicity, we presume the censor's blacklist starts with a single signature, which is a threshold on the value that a single feature takes on.

\paragraph{Polymorphism}

\begin{figure}%
\begin{center}
\includegraphics[width=0.7\linewidth]{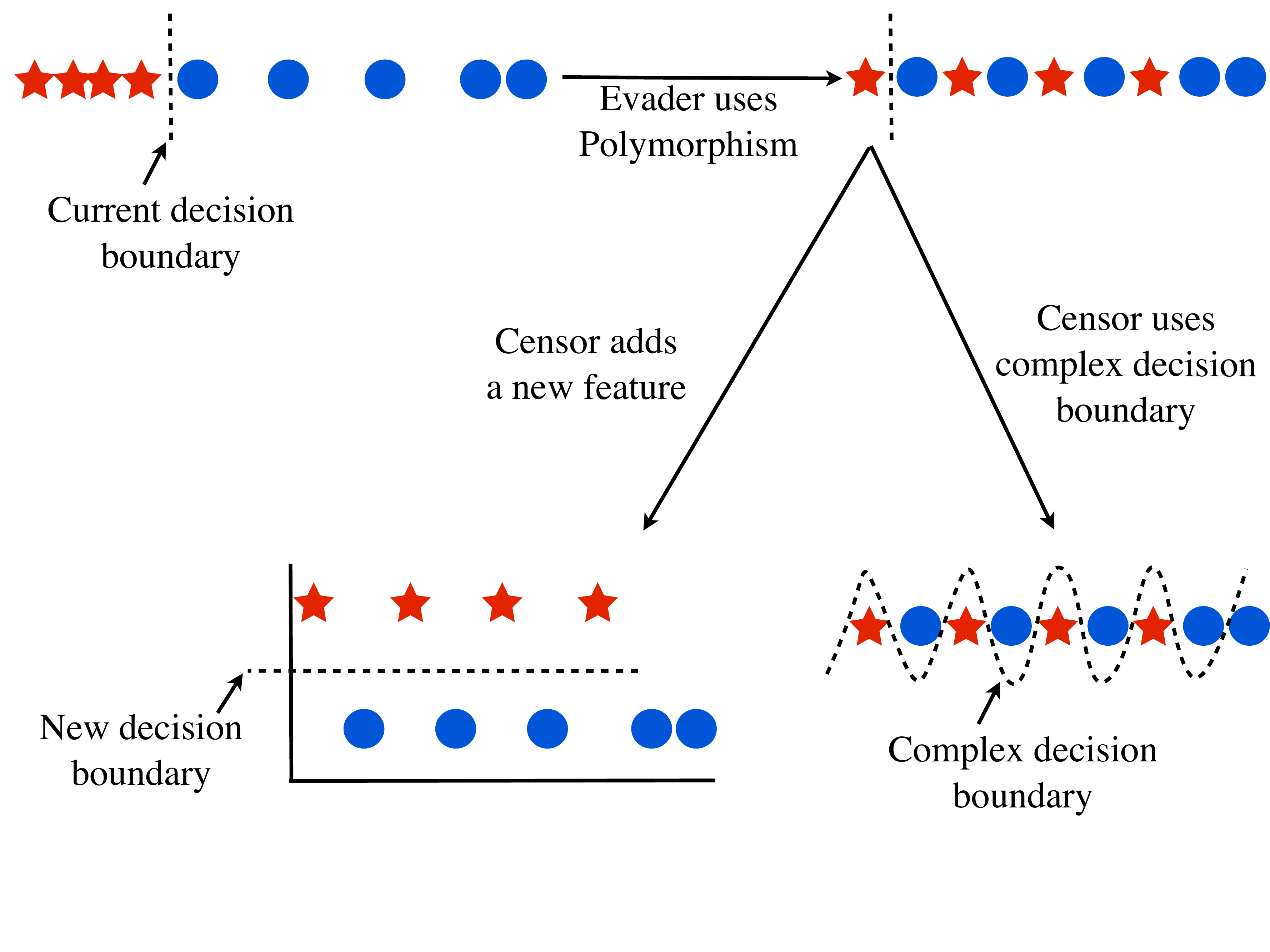}
\vspace{-10pt}
\caption{Polymorphism.  Red stars represent disallowed traffic and blue circles represent allowed traffic.  The dotted line shows the decision boundary of censor's classifier.}
 \label{fig:poly}
\end{center}
\vspace{-20pt}
\end{figure}

Polymorphism is a way of spreading out behavior.  To be polymorphic in a feature means that the feature takes on multiple values among different instances, such as messages.  
Spreading out the values of a feature used in a blacklist's signature can result in the signature no longer identifying disallowed traffic, increasing false negatives (Figure \ref{fig:poly}).  

In response, the censor can either come up with a new decision boundary using the old feature or employ a new feature.  A new boundary is likely to be more complex than the old one, making it more difficult to implement and less likely to generalize to new traffic (e.g.,~\cite{mitchell97book}).   Thus, it will typically be less accurate, increasing the costs of the censor.  Using a new feature may require additional measurements driving up the operating costs of the censor.  Either way, the censor must spend on the development effort. For example, many censors (Iran, China, and Syria) were blocking SSL and, thus, Tor as it also uses SSL. To circumvent this, Obfsproxy (obfs2) was implemented that added an extra layer of encryption on top of Tor with no recognizable byte patterns. To identify Tor with Obfsproxy the censor either has to employ a new complex classifier or a new feature. The new feature could be a passive feature like packet arrival time or an active feature like the reaction to active manipulation~\cite{houmansadr13sp}.

Note that the role of polymorphism in our description differs from that of Tor's pluggable transports.  The pluggable transports framework allows creating different protocols to transform the Tor traffic flow into different formats with the goal of replacing a blocked protocol quickly.  Here, we are envisioning a single protocol that automatically uses many polymorphic variants simultaneously. 

\paragraph{Steganography}

\begin{figure}%
\begin{center}
\includegraphics[width=0.7\linewidth]{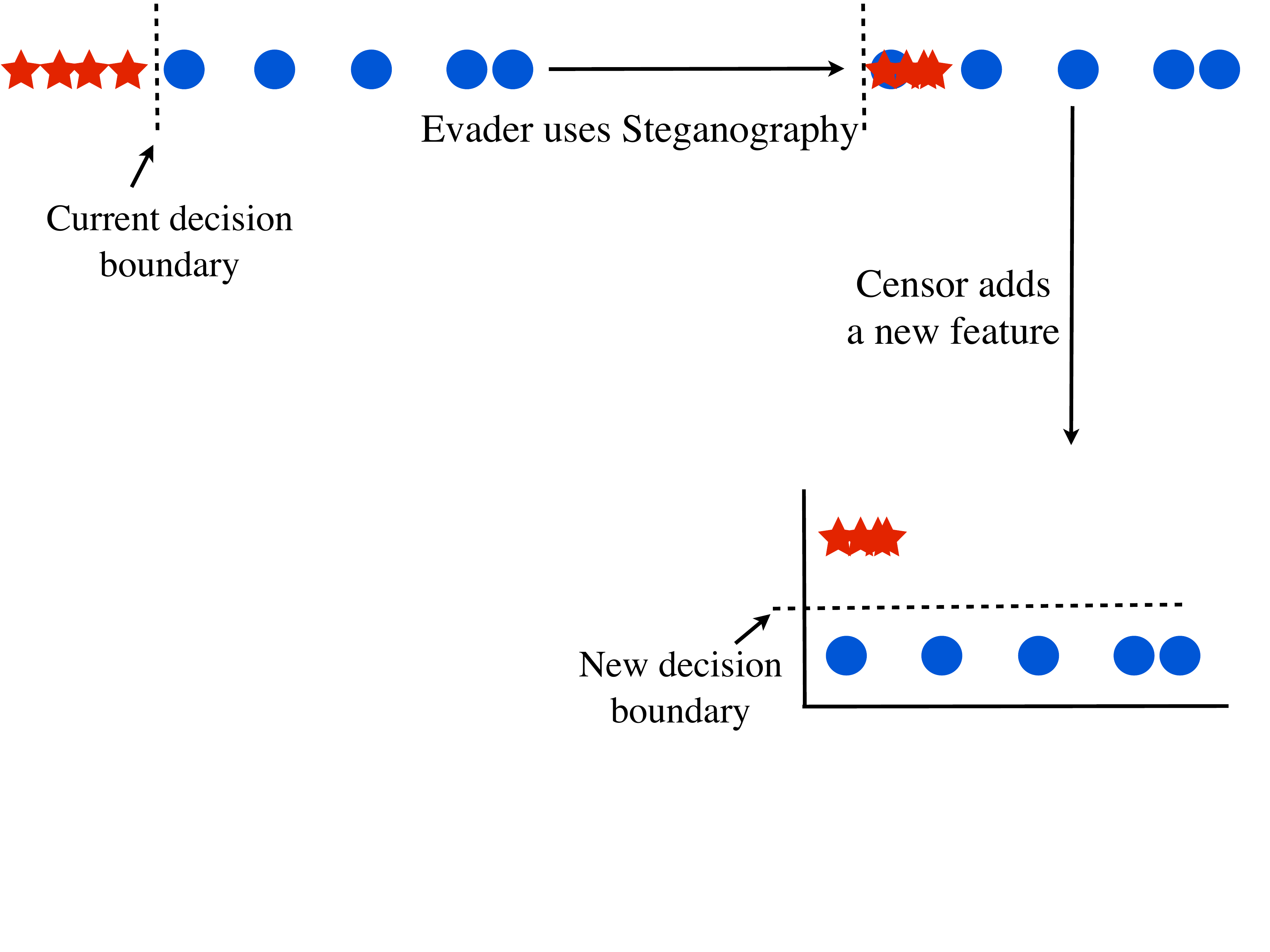}
\mbox{}\vspace{-30pt}\mbox{}
 \caption{Steganography}
 \label{fig:steg}
\end{center}
\vspace{-20pt}
\end{figure}

Steganography is a way of looking like allowed communications. To be steganographic in a feature means having values that are very close to the allowed communications.  Since steganography transforms a feature, as with polymorphism, it can result in unrecognizable disallowed traffic and more false negatives (Figure \ref{fig:steg}).

In the case of perfect steganography on the feature, responding by selecting a more complex decision boundary will not help since the traffic is no longer separable by the altered feature.
If the censor keeps relying upon the altered feature, it has to choose between raising false negatives or positives.

Alternatively, the censor can add a new feature on which allowed and disallowed traffic continue to differ, incurring the cost of implementing and tracking the new feature.  
Furthermore, adding such a new feature would require learning about how the disallowed and allowed messages differ, which could involve detailed knowledge of the protocol used by the allowed messages and approximated by disallowed traffic (the \emph{cover protocol}).  For example, SkypeMorph transforms Tor traffic to look similar to Skype traffic by changing the packet length distribution of Tor traffic.   To distinguish real Skype and mimicked Skype traffic, the censor can check their error behaviors using knowledge of the Skype protocol~\cite{houmansadr13sp}. 

\section{The Costs of Censorship}

Motivated by the above examples, we believe that censorship and censorship evasion has been and will continue to be an arms race where censors will track an increasing number of features and evaders will transform an increasing number of them.
Since the pace of this race is set largely  by the budgets of each opponent, technologies that increase the costs of the censor while not increasing the costs of the evaders should be welcomed by the evaders.  
For this reason, we present a simple cost model of the censor.  

First, we consider the costs that the censor incurs as a consequence of allowing or disallowing various packets.  
Let $c(t,a)$ be the cost to the censor when taking the action $a$ on a packet of type $t$ where for simplicity the actions are $a_{\msf{a}}$ for allow and $a_{\msf{d}}$ for disallow and the types are $t_{\msf{a}}$ for allowed or $t_{\msf{d}}$ for disallowed.  (Refinements are possible.)
The censor will pick the action that minimizes its expected cost given the information it has on the packet $i$.  That is, it selects
\begin{align}
d(i \given F) &= \argmin_{a} \sum_{t \in T} P(\msf{type}(i){=}t \given F(i)) * c(t, a) \label{eqn:d}
\end{align}
where $F(i)$ is the result of computing the features in the set $F$ on $i$, $T$ is the set of types, and $\msf{type}(i)$ is the type of $i$.
The cost that the censor will incur for $i$ is
\begin{align}
C(i \given F) &= c(\mathsf{type}(i), d(i \given F)) \label{eqn:C}
\end{align}

When $d(i \given F)$ is $a_{\msf{a}}$ but $\msf{type}(i)$ is $t_{\msf{d}}$, the censor has a false negative.  We assume that the cost $c(t_{\msf{d}}, a_{\msf{a}})$ is positive since the censor is attempting to block such messages.
When $d(i \given F)$ is $a_{\msf{d}}$ but $\msf{type}(i)$ is $t_{\msf{a}}$, the censor has a false positive.  We assume that $c(t_{\msf{a}}, a_{\msf{d}})$ is positive since blocking allowed traffic can disrupt the economy of the censoring country~\cite{robinson13oitp}.  The others costs could be zero or even negative, indicating a reward.

Second, we consider the costs of operating the censorship system.  
To compute the features in the set $F$, the system makes measurements of traffic, such as recording the packet arrival times or lengths.  
We model operating each measurement $m$ as incurring a cost $\op(m)$.

We also allow the censorship system to be stateful.
For example, state allows us to model the feature of flow entropy, which requires making measurements of each packet in a flow and storing distributional information for each flow.
The operating expenses of a feature due to storage can depend upon the accuracy to which it is computed.  For example, entropy-based approaches become more accurate as the number of samples increases.  However, sampling, in this case, means waiting for more packets, which means more storage costs.  (Also, if such packets are sent to their destinations before a decision is reached, this can lead to those packets being false negatives, also increasing cost.)  
With these considerations in mind, we conclude that typically, flow-level features computed over a distribution of packet-level features are typically more costly than packet-level features.
We model each feature $f$ as incurring a cost $\store(f)$ that depends upon the amount of memory it requires.

Lastly, we consider development costs.
Each new feature $f$ must be implemented, incurring a cost of $\imp(f)$.  

For simplicity, we assume that the censor's system is updated according to a development cycle of fixed duration.  To each cycle, we charge the costs of developing the classifier used during that cycle and for using it during the cycle.
The development costs depend upon not just the set $F'$ of features used during the cycle, but also the set $F$ of features used in the past since previously used features need not be reimplemented.  
We denote the expected total cost as $\cost(F' | F, P)$ where $F'$ are the features used by the classifier during the cycle, $F$ are the features previously used, and $P$ is the traffic distribution.
We model the expected total cost as
\begin{align}
\cost(F' | F, P) =  
\phantom{+}& \sum_{\vec{i}} P(\vec{i}) * \sum_{i \in \vec{i}} C(i | F')\\ 
+& \sum_{m \in \meas(F)} \op(m)\\ 
+& \sum_{f \in F'} \store(f)\\
+& \sum_{f \in F' - F} \imp(f)
\end{align}
where 
$\vec{i}$ ranges over the traffic possible during that cycle and
$\meas(F)$ denotes the set of measurements needed to compute the features $F$. 
The actual cost is computed by fixing $\vec{i}$ to the actual traffic.

We remind the reader that the above model is not designed to produce accurate predictions of the censor's actual costs.  Rather, we keep it simple and abstract while treating it as point of departure to discuss how to increase the censor's relative costs in general.  Thus, we make no effort to estimate any of the costs to which the model refers.  Furthermore, do not claim that our model captures every cost of the censor.  
With this model in mind, we now turn to the approaches that the evasion tool may use to drive up this cost.

\section{Evaluation Criteria}
\label{sec:crit}

Given that we cannot directly measure the costs effecting censors, we search for surrogates that protocol designers can measure and have reason to believe are correlated with these costs.  Such surrogates can serve as criteria for selecting which research proposals to deploy.  %
Since we do not empirically demonstrate that these surrogates actually correlate to a censor's costs, they must be considered hypotheses.

For a surrogate of the accuracy of a set of features, the tool evaluator may test a classifier using those features over simulated traffic that includes traffic from the evaluated tool.
The number of lines of code needed to implement a feature can act as a surrogate for the its implementation costs.
The amount of storage for each feature serves as a surrogate for storage costs.
The number of measurements needed for all the features is a surrogate for measurement costs.  

Intuitively, the quality of an evasion tool is proportional to the cost of the most inexpensive feature set $F$ that achieves such a level of quality as demanded by the censor.
Thus, if the evaluator can find an inexpensive set of features $F_1$ that accurately identifies traffic from a evasion tool $e_1$ 
but cannot find an equally or less expensive set of features $F_2$ that accurately identifies traffic from a tool $e_2$,
then the evaluator should suspect that $e_2$ is a better one than $e_1$. 
However, the evaluator must keep in mind that these findings are relative to his ability to find and test feature sets $F$ and to the method of estimating the expense of features.

In this way, the creators of tools cannot argue that their tool achieves some level success since some set of features might exist that they failed to consider.  However, they can demonstrate the shortcoming of other tools by illustrating feature sets under which other tools perform more poorly than their own.

Tool creators can also demonstrate that certain features, those obfuscated by their tool, are unlikely to be useful in crafting attacks against their own tool.  By doing so for inexpensive, well-known features, they can argue that either unusual or expensive features would have to be used for an attack against their system.  Thus, a heuristic metric of a tool's prospects for success is the cheapness of the features it obfuscates.  For example, a tool that obfuscates a few very inexpensive features should be viewed more favorably than a tool that obfuscates expensive features, even if it obfuscates more of them.

While computing these surrogates may appear daunting, those evaluating evasion tools typically implement and run programs using the features they claim their tool obfuscates.  Thus, they may use their own implementations to find values for these surrogates.

\section{Reevaluation of Existing Tools}
\label{sec:eval-tools}

In this section, we demonstrate how our evaluation criteria can be used in practice  to guide the development process and evaluation of different circumvention tools. To do this we reevaluate some existing tools using our evaluation criteria.
We examine only the capabilities of these tools as described in each paper's evaluation; we make no effort to confirm the correctness of these evaluations or to infer additional capabilities of the tools.  %
ScrambleSuit~\cite{winter2013scramblesuit} obfuscates Tor traffic 
by not employing a telltale TLS handshake used by plain Tor
and by polymorphically randomizing packet lengths and interpacket arrival times to look different from both Tor and its own instances.  
Presuming that the packet lengths and interpacket arrival times over which it randomizes matches ones seen in allowed traffic, then it obfuscates these features.

SkypeMorph~\cite{mohajeri2012skypemorph} is a steganography tool to obfuscate Tor traffic as Skype video calls. 
In addition to hiding the TLS handshake, it changes the Tor packet sizes and interpacket delays to match that of pre-recorded traffic of a Skype video call. 
Since Skype traffic is common, it obfuscates these two features.

StegoTorus is a polymorphic steganographic tool designed to have a diverse set of steganographic modules allowing the StegoTorus client to use whichever modules the censor has not yet blocked~\cite{weinberg2012stegotorus}.  Currently, there are two proof-of-concept steganography modules, one uses HTTP and another mimics an encrypted peer-to-peer cover protocol, such as Skype.
StegoTorus obfuscates the TLS handshake, connection length (seconds), connection payload, and per-packet payloads, which corresponds to two measurements since connection payload and per-packet payload require the same measurement.

While all three obfuscates a small number of inexpensive features, some of their effort to obfuscate distributional features may be misplaced.
For example, StegoTorus and SkypeMorph focus on mimicking distribution based features but fail to mimic inexpensive features like error codes, a weakness we discuss next.

\section{Active Manipulation}
\label{sec:active}

To illustrate the censor's ability to focus on inexpensive, simple features, we discuss active manipulation.
Active manipulation aims to increase $P(F(i) \given \msf{type}(i){=}t_{\msf{d}})$ when $\msf{type}(i) = t_{\msf{d}}$ by making the instance behave in a manner that is characteristic of disallowed traffic while not degrading any of the other probabilities.  To do so, active manipulation engages in behavior characteristic of an evasion tool that is not characteristic of any known allowed protocols.

For example, presuming that all Tor traffic is disallowed, the censor could initiate the Tor protocol handshake and observe whether the client producing the instance responds in the manner of Tor.  
Such manipulations work well since allowed traffic is unlikely to just so happen to exhibit Tor's complex behavior meaning that active manipulation introduces few false positives.
Thus, atypical, complex behaviors are dangerous for evasion tools as it provides a telltale sign of its use.
Systems like ScrambleSuit that reduces the complexity of the handshake for those without a password represent progress~\cite{winter2013scramblesuit}.

In Houmansadr et al.~\cite{houmansadr13sp}, they consider a form of active manipulation for defeating steganographic systems.
It operates in two steps.  First, it establishes that the protocol is very similar to some whitelisted protocol in a manner similar to the initiation attack discussed above.  Second, it proves that protocol is not really that whitelisted protocol by exercising some atypical behavior of the whilelisted protocol.  

Using some additional reasoning and replacing ``whitelisted traffic'' with ``allowed traffic'',  the above manipulation approach may also be used by a blacklisting censor.  
Intuitively, the above two steps together show that the protocol is masquerading as another, which is suspicious enough to warrant blocking since it is unlikely that allowed traffic would do so.
However, in this case, the evader may respond as in the above attacks by decreasing its complexity to pass as a simple, unknown protocol, which in this case would mean no longer attempting to look like a known protocol.  Thus, it appears that \emph{polymorphism is a better choice than steganography in the blacklisting case.}

\section{Discussion and Future Work}

The simple cost model presented in this preliminary proposal has already provided some insights into tool evaluation.  We have provided a proof of concept that evaluations can be more universally applied and more closely tied to economics than in the past. 

These advantages come with difficulties making our evaluations more complex than tool-specific ones with a narrow focus.  Furthermore, this preliminary work leaves some questions open.  We consider some below.

(1) Can our cost model be validated?
We developed our cost model primarily through intuition.  Ideally, we would have empirical results showing its accuracy.  Future work can look at the costs of actual censors in democratic countries with published budgets to gain insights into their costs.  We can also compare our model to the costs of those engaged in similar arms races, such as spam detection or network intrusion detection.

(2) How should we select features to examine?
The evaluator might overlook relatively cheap features while examining ones that the censor never intended to use.  In essence, the evaluations of StegoTorus and SkypeMorph, by not considering the features considered by Houmansadr et al.~\cite{houmansadr13sp}, suffered from this trap, which our methodology highlights but does not prevent.  
Future work can examine past data to estimate the relative cost of different features by seeing how long it takes for a censor to adapt to manipulations of them~\cite{khattak2013towards}.
However, the open-ended nature of the measurements and features possible makes it difficult to ever conclude that an evaluation examined every cheap feature.  We expect dealing with features elicited by manipulation, such as those used Houmansadr et al.\ to distinguish StegoTorus from its cover protocols~\cite{houmansadr13sp}, will be particularly difficult to characterize.  %

(3) Do our surrogates reflect the censor's actual costs?  
Answering this question depends upon the answers to the previous two: our surrogates must be developed from an accurate cost model and should focus on the features mostly used by the censor.
As a stopgap, our evaluation heuristic is computed using the features manipulated by the evasion tool rather than those used by the censor.  
Future work can instead subject evasion tools to a battery of tests based on inexpensive features.

(4) Are all costs equally important for any censorship regime? Different kind of costs might dominate in different censorship regimes. For example, China seems to increase censorship during and before specific events like the anniversary of Tiananmen Square protests, which suggests that the relative costs of false positives and negatives varies with the conditions.  Other countries (e.g. Saudi Arabia, Qatar) use computationally-costly DPI methods rather than using simpler but customized methods like DNS redirection, as China does~\cite{gillcharacterizing}.  These countries can easily buy off-the-shelf DPI tools but lack the skills to build customized tools, which implies that implementation costs might vary by country.

(5) What should we add to our model?  Our current model only considers features of packets.  Thus, we cannot fully evaluate tools like Flash Proxy~\cite{fifield2012evading} as it focuses on lowering the cost of creating new proxies rather than on obfuscating features of traffic.
We also assume that packets do not interact with each other, which limits our ability to explain attacks leveraging such interactions.
For example, Geddes et al.~\cite{geddes2013cover} show that randomly dropping $5\%$ of the traffic would render disallowed traffic using Skype as a cover protocol (e.g., FreeWave~\cite{houmansadr2013want} and SkypeMorph~\cite{mohajeri2012skypemorph}) useless without affecting legitimate Skype.  This is caused by a channel mismatch between the cover protocol and the disallowed protocol: the cover protocols are loss tolerant peer-to-peer systems whereas the disallowed protocol (Tor) is a loss intolerant client-proxy system.  We cannot model loss tolerance since we consider each packet in isolation.

(6) What are the trade-offs between the censor's and evaders' costs?  
Evaluations of evasion tools must also consider the costs of the evaders.  
Since evaders separate out into tool developers and tool users, we may require two additional cost models.  With these models in hand, we may examine the trade-off between increasing the censor's costs and keeping the evaders' costs acceptable.  
Unfortunately, fully understanding these trade-offs may require models that produce quantitative predictions of costs allowing us to compare each party's cost on a common scale.

In particular, by focusing on the censor's costs, we presuppose that the evaders are better off whenever the censor is worse off.  However, this is not always the case.  For example, an evasion technique that increases the censor's false positives but does not result in more disallowed traffic flowing makes the censor worse off without helping the evader.  Alternatively, the censor changing its opinion on the dangers of disallowed traffic may result in it blocking less disallowed traffic but without its costs increasing.  Thus, we would like a better understanding of when increasing a censor's costs corresponds to an improvement for the evader.

While this proposal does not answer these questions, our overall methodology allows us to systematically consider them. It makes plain that prior evaluations and our own criteria found in Section~\ref{sec:crit} must be treated as heuristics and encourages the evaluator to consider how they may deviate from the actual quantity of interest, costs. We provide a rigorous methodology in which to discuss the trade-offs among evasion tools and their evaluations.

\section*{Acknowledgments}
We thank David Fifield for many helpful conversations on this topic.

\balance

\bibliographystyle{abbrv}

\bibliography{censor}

\end{document}